\documentclass[a4paper,11pt]{article}
\usepackage{amssymb}
\usepackage{mathrsfs}
\usepackage{graphicx}
\begin{document}
\begin{center}
\LARGE{Invariant lengths using existing Special Relativity}
\end{center}
\begin{center}
Christopher D. Burton
\end{center}
\begin{center}
\textsl{George and Cynthia Woods Mitchell Institute for Fundamental Physics and\\
Astronomy, Texas A$\&$M University, College Station, TX 77843, USA}
\end{center}
\begin{center}
\textsl{email: chris.burton@tamu.edu}
\end{center}
\
\
\begin{center}
ABSTRACT
\end{center}
A field of random space-time events exhibiting complete spatial-temporal randomness appears statistically identical to all observers.  Boost invariant lengths naturally emerge when we examine fluctuation scales of this field such as the nearest neighbor distance.  If we interpret Planck's length as the characteristic fluctuation scale of quantum gravity, its boost invariance can then be understood without modifying Special Relativity.
\newpage
\begin{center}
\LARGE{\textsc{Introduction}}
\end{center}
\begin{sloppypar}
	An outstanding problem of special relativity is the invariance of Planck's length.  How can a length be the same for all observers?  The methods of ``Double Special Relativity'' introduced by Amelino-Camelia [1] and furthered by Magueijo and Smolin [2] relied upon deformed energy-momentum dispersion relations in the former and a modified generator of momentum space boosts in the latter to produce an invariant length or energy respectively.  These techniques were later found by Jafari and Shariati [3] to be re-descriptions of special relativity in non-conventional coordinates.
\end{sloppypar}
\begin{sloppypar}
	A different approach to this problem would be to examine the appearance of random events for different observers.  An example would be to consider lightning strikes over an area during a prescribed interval of time.  Assuming these lightning strikes exhibit complete spatial randomness, one observer could establish the probability of $n$ strikes per area.  Another observer in a rotated frame could do the same and would find the same probability.  The first observer could calculate various statistical quantities such as the nearest neighbor distance.  In doing so he would obtain a length $L$ which is the expected radius to the nearest lightning strike.  The second observer could do the same and would obtain the same length.  These two observers could then proclaim their discovery of a vector which remains invariant under rotations. Their error being that $L$ is not the component of a vector but rather it is the characteristic length of a fluctuation scale which is invariant under rotations.  
\end{sloppypar}
\begin{sloppypar} 
We now examine this same experiment with random spacetime events under the Lorenz transformation.
\end{sloppypar}
\begin{center}
\LARGE{\textsc{Random Events}}
\end{center}
\begin{sloppypar}
Since it is not possible to pinpoint the rest frame of a spontaneous random event, it will not be possible to pinpoint the rest frame of a collection of such events exhibiting complete spatial-temporal randomness.  The appearance and effects of  these random events will then be the same for all observers.  By definition, these events are independent and have equal a-priori probability to occur at any place and at any time.  Working in $3$ space dimensions with a Minkowski metric, observers could compare data by choosing, in their own rest frame, an agreed upon space volume of an agreed upon shape and count events over an agreed upon time interval.  For example, an $L \times L \times L$ cube observed for a time interval $T$.  Denoting this 4-volume sample by $V$, let $P_{V}(N)$ be the probability of observing $N$ events in $V$.  We can partition this 4-volume $V$ into $S$ identically shaped regions each with 4-volume $w$. Our probability would then be
\begin{eqnarray}
P_{V}(N) = \sum_{s_{0},s_{1},\ldots}{S\choose s_{0} \ s_{1} \ldots} [P_{w}(0)]^{s_{0}}[P_{w}(1)]^{s_{1}}[P_{w}(2)]^{s_{2}} \ldots
\end{eqnarray}
where $s_{0}$ is the number of samples with 0 events, $s_{1}$ is the number of samples with 1 event, etc., with $ V = S \cdot w,\  S = s_{0} + s_{0} + s_{0} + \ldots$ and $ N = 0 \cdot s_{0} + 1 \cdot s_{1} + 2 \cdot s_{2} + \ldots$
\end{sloppypar}
\begin{sloppypar}
Due to the properties of random events, we are free to relocate these $S$ regions arbitrarily which will not change $P_{V}(N)$  By making S arbitrarily large and $w$ arbitrarily small while keeping their product $V$ constant, we can deform $V$ into an arbitrary shape.  The resulting probability $P_{V}(N)$ is therefore independent of its shape.  Since the Lorentz transformation preserves 4-volume, $P_{V}(N)$ is invariant.  Since $P_{V}(N)$ is invariant, any statistical test of complete spatial-temporal randomness will yield the same result for all observers.  A concrete example would be the homogeneous Poisson Process with:
\begin{eqnarray}
P_{V}(N) = \frac {e^{- \lambda V}(\lambda V)^{N}}{N!}
\end{eqnarray}
\end{sloppypar}
\begin{sloppypar}
In order to explicitly demonstrate that random events transform into random events with identical statistical properties under a boost, the coordinate system $\epsilon = \frac{(t+x)}{\sqrt{2}}, \eta = \frac{(t-x)}{\sqrt{2}}$ will be used.  To compare data, observers would agree upon a sample size and shape observed from their own rest frame.  We will use an $L \times L$ square centered on the origin with edges parallel to the coordinate axes.  If we now perform a boost on the border of the square only (and not on the background of random events nor on the coordinate system), our square is transformed into a rectangle of dimensions $L'_{\epsilon} = e^{-\phi}L$ and $L'_{\eta} = e^{\phi}L$.  Since the volume of our rectangle remains unchanged, its probability distribution remains unchanged.  But this rectangle as viewed in the $(\epsilon, \eta)$ system is an $L \times L$ square in the rest frame of an observer moving with velocity parameter $\phi$ relative to the $(\epsilon, \eta)$ system.  Thus, when viewing any $L \times L$ square within their own rest frame, all observers agree upon $P_{V}(N), \overline{N}, \overline{\rho} = \frac{\overline{N}}{L^{2}}$, and all higher statistical moments of these quantities.  
\end{sloppypar}
\begin{sloppypar}
If we consider the active viewpoint and transform the events, intuitively one expects them to aggregate along lines or form correlations.  This is not the case.  Consider an $L \times L$ box as described above.  Partition this box into an arbitrarily large number $m$ of identical thin strips parallel to either the $\epsilon$ or $\eta$ directions.   Each thin strip has the same probability distribution function.  We now perform a boost on the box and the strips only (and not on the background of random events).  The probability distribution function of each strip remains unchanged.  If we now perform the reverse boost on the box, the strips, and on the background of random events, we will be able to ``see'' what the appropriately moving observer ``saw''. Since the probability distribution function of each strip again remains unchanged, there will be neither aggregations nor correlations amongst the random events. 
\end{sloppypar}
\begin{sloppypar}
This outcome is to be expected.  Random events have no correlations and thus have minimal information.  A boost cannot add (or remove) information to (or from) these events
\end{sloppypar}
Finally, let's examine the position moments of the events:				 			
\begin{eqnarray}
M^{ab} \equiv \frac{1}{N} \sum_{k=1}^{N}(\epsilon^{a}\eta^{b})_{k}
\end{eqnarray}
where we sum the positions of the $N$ events labeled by $k$ which fall within the $L \times L$ square.  To find the average $\overline{M^{ab}}$, we write $N$ as $\overline{N} \lambda$ where $\overline{N}$ is the average number of events in a single square: 
\begin{eqnarray}
\overline{M^{ab}} = \lim_{\lambda\to\infty}\frac{1}{\overline{N}\lambda }\sum_{\lambda = 1}^{\overline{N}\lambda}(\epsilon^{a}\eta^{b})_{k}
\end{eqnarray}
Partitioning the square into small elements, the sum becomes an integral which yields:    
\begin{eqnarray}
\overline{M^{ab}} = \Big(\frac{1}{4}\Big)\Big(\frac{1}{a+1}\Big)\Big(\frac{1}{b+1}\Big)\Big(\frac{L}{2}\Big)^{a+b}[1+(-1)^{a}][1+(-1)^{b}]
\end{eqnarray}
Since all observers use the same agreed upon $L$ within their own boxes, all position moments are invariant.  
To show that all statistical moments of $M^{ab}$ are invariant, we define
\begin{eqnarray}
(M^{ab})^{p} \equiv M^{(ap)(bp)} \equiv \frac{1}{N}\sum_{k=1}^{N}(\epsilon^{ap}\eta^{bp})_{k}
\end{eqnarray}
And form the multinomial:
\begin{eqnarray}
( M^{ab} - \overline{M^{ab}} )^{p} = \sum_{k=0}^{p}{p \choose k} [M^{ab}]^{p-k} [\overline{M^{ab}}]^{k} (-1)^{k}
\end{eqnarray}
Taking the average of this multinomial, we note that its terms are of the form $\overline{M^{cd}}(\overline{M^{ef}})^{g}$ which are manifestly invariant from previous arguments.  
\begin{center}
\LARGE{\textsc{Length Scales}}
\end{center}
\begin{sloppypar}
A field of random events has a characteristic scale.  For example, given a random process in Minkowski spacetime, we could ask the nearest neighbor question:  At what length $L$ would we expect to find zero events lying within the 4-volume defined by:
\begin{eqnarray}
x_{0}^{2} + x_{1}^{2} + x_{2}^{2} + x_{3}^{2} = L^{2}
\end{eqnarray}
This length $L$ will be the same for all observers.  
\end{sloppypar}
\begin{sloppypar}
This idea can be extended to curvature fluctuations.  In principle we could devise an experiment measuring the value of the Ricci scalar at different locations.  Focusing on a Minkowski spacetime with no matter sources, the classic field equations tell us we would always measure zero.  However, we know there are quantum fluctuations of the curvature.  A fully functioning theory of quantum gravity would give us the expectation value of the observable $R$.  Because of fluctuations, we would expect $<R>=0$ but a non-zero variance $<R^{2}>$.  Since $R$ is an invariant, all statistical moments of $R$ and the probability distribution of $R$ must be invariant otherwise we could pick out a preferred reference frame.  The boost invariant fluctuation scale $\frac{1}{<R^{2}>^{1/4}}$ has the units of length, an infinite value when there are no quantum fluctuations, and can be thought of as a measure of the local "bumpiness" of this geometry.  It then seems reasonable to associate Planck's length with an invariant quantum fluctuation scale, such as ($\frac{1}{\sigma_{R}})^{1/2}$.
\end{sloppypar}
\begin{center}
\LARGE{\textsc{Conclusion}}
\end{center}
\begin{sloppypar}
Boost invariant lengths naturally emerge when we consider random events and their transformations.  A field of random events transforms into a field of random events with identical statistical properties.  The observer independence of Planck's length, when interpreted as the characteristic fluctuation scale of quantum gravity, can then be understood in these terms.
\end{sloppypar}
\begin{flushleft}
\Large{\textsc{References}}
\end{flushleft}
$[1]$G. Amelino-Camelia, Int. J. Mod. Phys. D 11, 1643 (2002) (arXiv:gr-qc/0210063)\\
$[2]$J. Magueijo and L. Smolin, Phys. Rev. D 67, 044017 (2003) (arXiv:gr-qc/0207085)\\
$[3]$N. Jafari and A. Shariati, AIP Conf. Proc. 841, 462 (2006) (arXiv:gr-qc/0602075)
\end{document}